\documentclass[9pt,twocolumn,twoside]{revtex4-2}
\usepackage{graphicx}
\usepackage{amsmath}

\usepackage{fancyhdr}

\begin{document}

\title{\fontsize{12pt}{10pt}\selectfont
Shear-induced mechanical yielding controls cluster morphology and dynamics \\ in aggregating non-Brownian suspensions}

% Use letters for affiliations, numbers to show equal authorship (if applicable) and to indicate the corresponding author
\author{Jae Hwan Jeong}
\author{Donald L. Koch}
\author{Sarah Hormozi}

\affiliation{Robert Frederick Smith School of Chemical and Biomolecular Engineering, Cornell University, Ithaca,
NY 14853, USA}

\begin{abstract}
\noindent In sheared aggregating suspensions, flow drives both the formation and breakage of particle clusters, whose dynamic evolution influences transport and rheological properties. Understanding these mechanisms remains challenging due to the complex interplay between interparticle forces, hydrodynamic interactions, and shear flow. Using parallelized Accelerated Stokesian Dynamics (PASD) simulations, we investigate cluster microstructure and rotational dynamics, and how their coupling produces suspension stress in sheared, attractive non-Brownian suspensions. When interparticle attraction dominates over separating shear forces, clusters grow until mechanical loading balances maximum interparticle bond strength. We examine how the cluster stresslet depends on cluster size, with both quantities normalized by their mechanical yielding scales. Scaled cluster stresslets collapse onto a master curve exhibiting three distinct regimes: cubic, quadratic, and linear with respect to the scaled radius of gyration $R_g$. In the cubic regime, small, compact clusters behave as rigid bodies, whose hydrodynamic stresslet scales with cluster volume $R_g^3$. As clusters grow into the quadratic regime, low-aspect-ratio clusters preferentially yield under flow, driving a proportional increase in average aspect ratio with size. High-aspect-ratio clusters follow Jeffery rotational dynamics within the flow-gradient plane, spending longer durations aligned with the flow and thereby reducing their time-averaged hydrodynamic stresslet to scale as $R_g^2$. In the linear regime, clusters approach the maximum size sustainable by the interparticle bonds. The internal tension is bounded by the maximum bond strength, limiting the stresslet to an $R_g$ dependence. These findings establish a mechanistic framework connecting microscale cluster morphology and dynamics to macroscale rheology.
\end{abstract}

\maketitle
\thispagestyle{fancy}

\noindent Aggregating suspensions exhibit complex phenomena under flow, as evolving cluster size and morphology affect suspension properties such as viscosity and conductivity\ \cite{papadopoulou2020shear, hipp2024ElecTransport}. These dynamic responses are important in a wide range of industrial applications. In semi-solid flow batteries\ \cite{duduta2011SSFB, fan2014polysulfideSSFB}, for instance, electrochemically active particles and conductive additives form a percolating gel network at rest. As the suspension is pumped through the electrochemical cell, this network yields and fragments into clusters, whose subsequent evolution influences the fluid's transport and electrical efficiencies\ \cite{brunini2012modelingSSFB, li2013SSFB}. In water deionization, activated carbon (AC) particles with low surface charge aggregate rapidly into larger clusters, facilitating charge transport and improving desalination performance\ \cite{das2024surfactantACparticles, jin2025ACparticles}. Polymer-bridging flocculation provides another broad example spanning pharmaceuticals, food, and cosmetics, where flow-induced cluster collision and breakup are engineered to produce targeted-size clusters\ \cite{biggs2000PolymerBridging, pongjanyakul2009Pharma, agarwal2009PolymerBridgingModel}. In such systems, primary particles ranging from tens of nanometers to tens of micrometers form large clusters via strong interparticle attractions. Suspending these large clusters against sedimentation requires a high-viscosity fluid and high shear rates. These conditions cause hydrodynamics to dominate over Brownian motion at the cluster scale, allowing the aggregated particles to behave as non-Brownian clusters\ \cite{Zaccone2011BrowntoNonBrown}. In this regime, the competition between interparticle attraction and hydrodynamics controls cluster formation and breakup, thereby influencing suspension properties. Thus, understanding these mechanisms is essential for process design.
%\Endparasplit

Although numerical simulations provide a bridge between microscopic interactions, cluster formation, and suspension rheology, resolving the combined effects remains challenging. Applied flow induces particles to form clusters while simultaneously reorienting, deforming, and rupturing them. Microscale attractions drive particle aggregation, while frictional contacts at particle surfaces stabilize interparticle bonds. These interactions disturb the surrounding fluid, which in turn modifies both particle dynamics and hydrodynamic stresses. Recent studies have explored some of these aspects. For example, Varga et al.\ \cite{VargaSwan2019PatternFormation} revealed that the long-ranged hydrodynamic interactions modeled by the Rotne-Prager Yamakawa (RPY) tensor capture hydrodynamic influence on the formation of vorticity-aligned anisotropic clusters. However, models built with the RPY tensor for hydrodynamics neglect both lubrication and many-body interactions because the tensor itself captures only leading-order pair interactions at wide separations\ \cite{swan2016rapid}. This can impart significant changes in the rate of compaction or stretching of interparticle bonds and, consequently, the morphology and dynamics of particle strands within clusters\ \cite{Zia2017pairMobility, varga2018large}. Other efforts have explored the roles of friction\ \cite{manounou2014discrete}, non-spherical shapes\ \cite{Irani2014AggregatingDisks}, and lubrication interactions\ \cite{Singh2019YiedlingtoJamming} in coupling cluster structure to rheological properties. But, the lack of long-ranged many-body hydrodynamic interactions in their models imposes a significant limit on the exploration of the impact of hydrodynamics on transport properties such as high-frequency viscosity\ \cite{abade2010HighFreqVisc} and collision-driven diffusivity\ \cite{cichocki1999ThreeBody, lin2022diffusivity}. 

\begin{figure*}[tbhp]
\centering
\includegraphics[width=\linewidth]{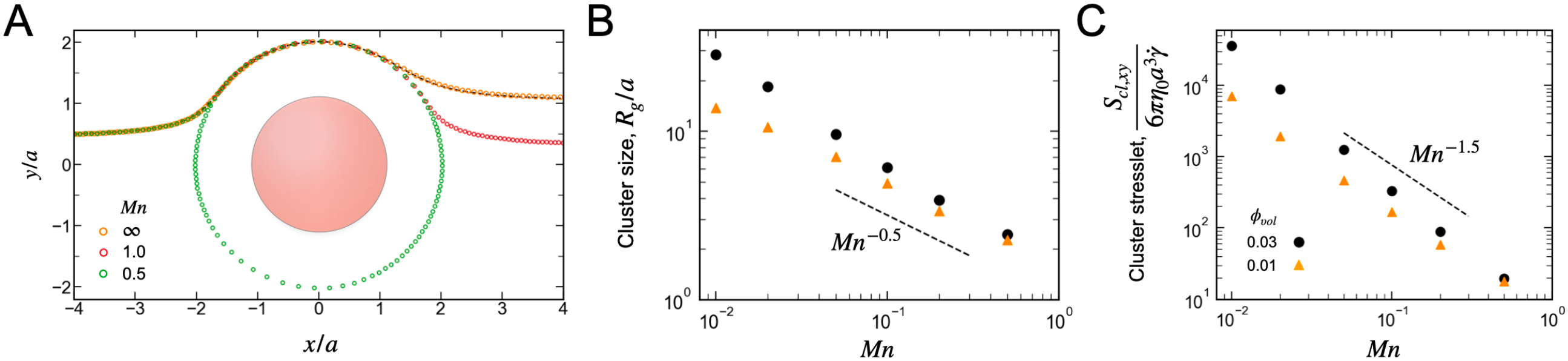}
\caption{Effect of $\mathit{Mn}$ on pair-particle trajectories and the size and stresslet of clusters in sheared suspensions. (A) Relative trajectories of a pair of particles for different $\mathit{Mn}$. Symbols represent simulation results, and the dashed line represents a theoretical trajectory given by Da Cunha and Hinch\ \cite{da1996CunhaHinch}. (B) Size $R_g$ and (C) stresslet $S_{cl,xy}$ of clusters as a function of $\mathit{Mn}$ for volume fractions $\phi_{\mathit{vol}} = 0.01$ and $0.03$. $R_g$ results represent stresslet-weighted averages, i.e., $\sum_i S_{cl,xy,i}R_{g,i} \big/ \sum_i S_{cl,xy,i}$ and $S_{\mathit{cl,xy}}$ results represent volume-weighted averages, $\sum_i R^3_{g,i}S_{cl,xy,i} \big/ \sum_i R^3_{g,i}$, where the sum over $i$ accounts for all clusters in each simulation snapshot collected over 200 strains at steady state.}
\label{fig:1}
\end{figure*}

To address this challenge, we simulate sheared aggregating suspensions of hydrodynamically coupled particles using parallelized Accelerated Stokesian dynamics (PASD) simulations\ \cite{fiore2018rapid, fiore2019FSD}. Building on recent GPU-based frameworks\ \cite{fiore2019FSD}, our approach utilizes the linearity of Stokes flow by decomposing near-field (which includes lubrication forces) and far-field many-body hydrodynamic contributions\ \cite{brady1988stokesian, sierou2001accelerated, wang2016spectral} and applying distinct parallelization strategies to each. We extend this framework to incorporate attractive and frictional interactions, optimizing it for a near-field GPU-parallelized scheme. This reduces the computational complexity to $O(N_p)$\ \cite{fiore2019FSD}, where $N_p$ is the number of particles, enabling large-scale simulations. 

In this work, we show how the interplay between interparticle attraction and hydrodynamics affects cluster microstructure and dynamics, which in turn produces the rheology of sheared aggregating non-Brownian suspensions. The competition between the separating shear force and the attractive force, $F^A$, is characterized by the Mason number, $\mathit{Mn} = 6\pi\eta_0 a^2\dot{\gamma}/F^A$, where $\eta_0$ is the fluid viscosity, $a$ is the particle radius, $\dot{\gamma}$ is the imposed shear rate, and we focus on the regime in which attractions remain dominant $(\mathit{Mn}<1)$. In this regime, shear-induced mechanical forces are insufficient to rupture interparticle bonds within small clusters, allowing clusters to grow until the mechanical force formed across the cluster overcomes the maximum bond loading. This force balance produces a scaling law for cluster size as a function of $\mathit{Mn}$ and the suspension volume fraction, $\phi_{\mathit{vol}}$. We further show that cluster shape and orientational dynamics influence this loading and the resulting cluster stresslets. These relationships collapse onto unified scaling laws that connect cluster size, shape, and orientations to stresslets, which demonstrate the underlying yielding mechanisms in sheared aggregating suspensions.
\section*{Numerical Methods}
\noindent We study non-Brownian suspensions in Stokes flow that interact via hydrodynamic $F^H$, attractive $F^A$, and contact $F^C$ forces. Rigid spheres of radius $a$ are suspended in an incompressible Newtonian fluid of viscosity $\eta_0$ in a cubic box with length $L$, with Lees–Edwards boundary conditions. We explore dilute particle volume fractions, $\phi_{\mathit{vol}} = 0.01$ and $0.03$, which allow analysis of individual cluster rotation and yielding dynamics prior to subsequent collision events. To prevent finite-size effects, the number of particles $N_p$ and the corresponding box size are chosen such that $L/2>2\langle R_g\rangle_S$, where $\langle R_g\rangle_S$ is the stresslet-weighted average cluster radius of gyration defined in Fig.~\ref{fig:1}B. The minimum $N_p$ is 8,000, and it increases to 27,000 for the lowest $\mathit{Mn}$.

Interparticle attraction is modeled by a quadratic potential characterized by a well depth $U_{max}$ and a maximum bond length $\delta_{bond}=0.1a$. Particle contact occurs when surface asperities of height $h_r=0.01a$ overlap, at which point normal and tangential contact forces come into play\ \cite{andreotti2013granular}. For a pair of particles separated by a centre-to-centre distance $r$, we define the separation distance as $\delta=2(a+h_r)-r$, where $\delta>0$ denotes asperity overlap. The interparticle potential is given by
\begin{equation}
U(\delta)=
\begin{cases}
0, & \delta<-\delta_{bond},\\
-U_{max}\left(1-\dfrac{\delta^2}{\delta_{bond}^2}\right),
&-\delta_{bond}\leq\delta\leq0,\\
-U_{max}+\dfrac{2}{5}k_n\delta^{5/2},&\delta>0,
\end{cases}
\label{eq:methods_potential}
\end{equation}
where $k_n$ is the normal stiffness of the Hertzian spring\ \cite{gallier2014Friction, mari2015DST, zhang2024friction}. $k_n$ is chosen for each $\mathit{Mn}$ such that the average $\delta$ remains below $h_r$. The maximum attractive force occurs at $\delta=-\delta_{bond}$, with a magnitude of $F^{A}_{max}=2U_{max}/\delta_{bond}$. Defining $\mathit{Mn}$ relative to $F^{A}_{max}$ ensures that $\mathit{Mn}<1$ corresponds to regimes where the applied shear force is weaker than the maximum interparticle attraction for a single particle pair. Tangential contact forces are modeled using a linear spring, $\mathbf{F}_t^C=-k_t\boldsymbol{\ell}_t$, where $k_t$ is the tangential stiffness and $\boldsymbol{\ell}_t$ is the tangential displacement during contact. The tangential force is constrained by Coulomb’s friction law, $\left|\mathbf{F}_t^C\right|\leq\mu_d\left|\mathbf{F}_n^C\right|$, and we set $\mu_d=1$\ \cite{mari2015DST, fernandez2013Friction}.
\section*{Results}
\noindent We first consider a pair of particles to examine whether the Mason number $\mathit{Mn}$ correctly characterizes the competition between the separating shear force and interparticle attraction. In Fig.~\ref{fig:1}A, one particle is fixed, while the second moves according to the relative velocity arising from the imposed shear flow, hydrodynamic interactions, and interparticle forces. In the absence of attraction ($Mn = \infty$), the simulated trajectories follow the analytical solution of Da Cunha and Hinch for two rough spheres with a surface roughness $h_r$ in simple shear flow\ \cite{da1996CunhaHinch}. The trajectories are asymmetric because normal contact forces prevent the particles from approaching closer than $2h_r$, causing them to separate on streamlines farther apart than on their approach. When attractive forces are introduced and dominate over the separating shear force ($Mn < 1$), the particles remain bound and slide along each other’s surfaces. At $Mn = 1$, the shear force becomes comparable to the maximum attractive bond force, showing the transition between bound and separating trajectories. This two-particle problem confirms that $\mathit{Mn}$ captures the competition between the microscopic forces driving shear-induced separation and interparticle attraction.

The $\mathit{Mn}$ dependence of the cluster size $R_g$ and stresslet $S_{cl,xy}$ in sheared suspensions is shown in Fig.~\ref{fig:1}B and C for different $\mathit{Mn}$ and $\phi_{\mathit{vol}}$. At a given volume fraction, the cluster size and stresslet collapse onto linear master curves scaled by $\mathit{Mn}$. This suggests that $\mathit{Mn}$ accurately describes the relative magnitude of the mechanical rupturing force against the attractive force developed within the clusters. The $R_g$ exhibits an $Mn^{-0.5}$ scaling (Fig.~\ref{fig:1}B), and $S_{cl,xy}$ exhibits an $Mn^{-1.5}$ scaling (Fig.~\ref{fig:1}C). We note that the cluster stresslet scales as $R_g^3$, as substituting the $R_g$ with its scaling ($Mn^{-0.5}$) produces $Mn^{-1.5}$ scaling. This implies that particle clusters behave as rigid bodies, whose stresslet is proportional to their volume, i.e., $R^3$ in shear flow. Both $R_g$ and $S_{cl,xy}$ are larger at a given $\mathit{Mn}$ for the higher volume fraction. This can be attributed to the increased frequency of cluster-cluster and cluster-particle collisions at higher volume fractions, which allows clusters to grow larger before mechanically yielding. 

\begin{figure}[t!]
\centering
\includegraphics[width=\linewidth]{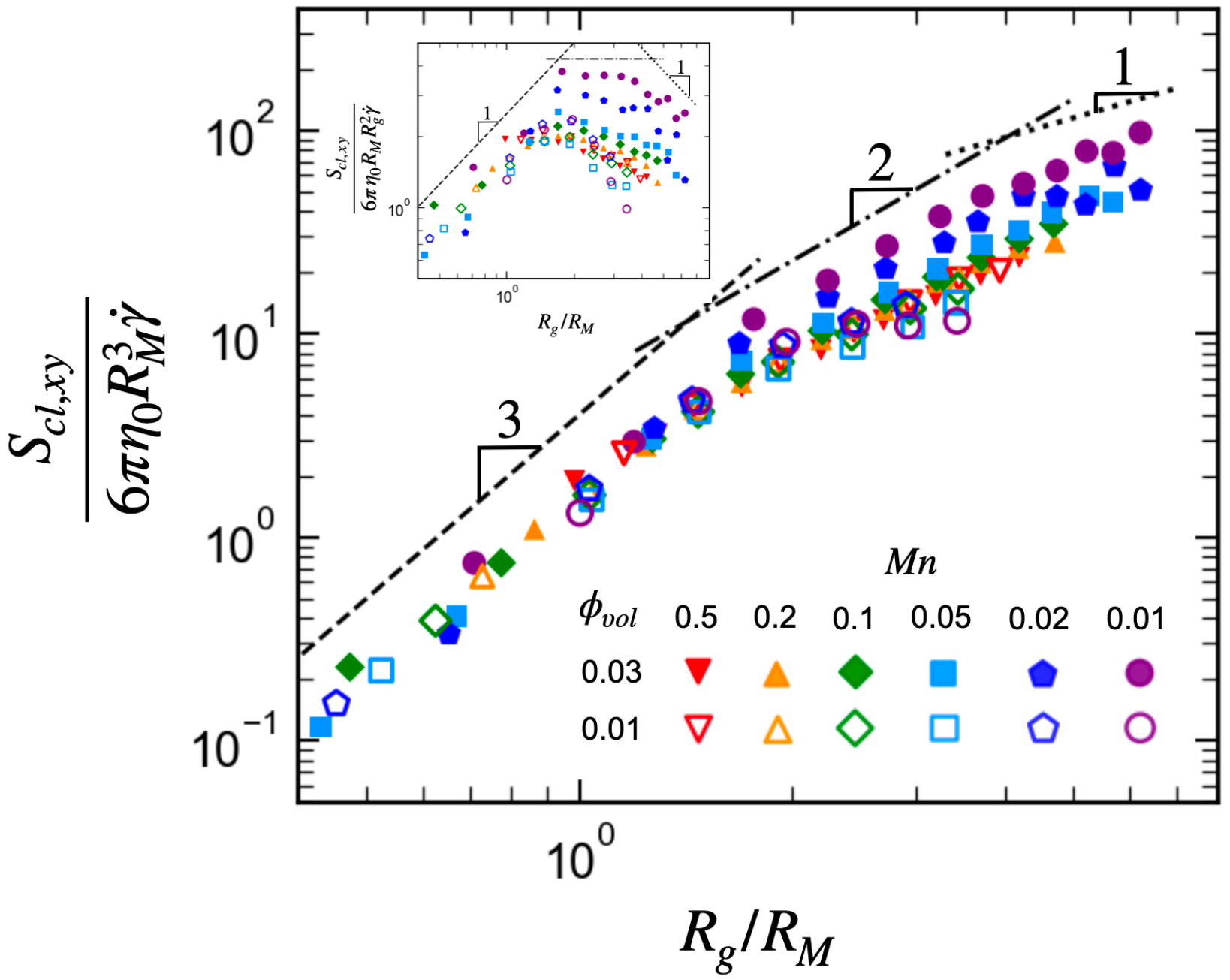}
\caption{
Collapse of cluster stresslets as a function of cluster radius of gyration for $\phi_{\mathit{vol}} = 0.01$ and $0.03$. The cluster radius $R_g$ and stresslet $S_{\mathit{cl,xy}}$ are scaled by the mechanical yielding scales $R_M$ and $6\pi \eta_0 R_M^3 \dot{\gamma}$, respectively. Results include clusters identified in simulation snapshots collected over 200 strains at steady state. Dashed guidelines indicate $R_g^3$, $R_g^2$, and $R_g$ scaling. The inset shows $S_{cl,xy}/(6\pi \eta_0 R_M R_g^2 \dot{\gamma})$ as a function of $R_g/R_M$, highlighting the intermediate $R_g^2$ regime as a plateau that extends to larger $R_g/R_M$ at higher $\phi_{\mathit{vol}}$ and lower $\mathit{Mn}$. Bins containing fewer than $0.01\%$ of the total sampled clusters for a given $\mathit{Mn}$ and $\phi_{\mathit{vol}}$ are omitted.
}
\label{fig:2}
\end{figure}

To understand the physical origin of the $Mn^{-0.5}$ scaling, we consider a cluster of size $R_g$ undergoing shear-induced mechanical yielding. The mechanical load on the rigid cluster exerted by the flow scales as $6\pi \eta_0 R_g^2 \dot{\gamma}$. We then assume that the applied load is transmitted through a single load-bearing bond connecting two parts of the cluster\ \cite{potanin1993aggregate}. The cluster will break when the applied force exceeds the maximum bond force, $2U_{max}/\delta_{bond}$. By equating these two forces, the critical cluster radius for mechanical yielding, $R_M$, is given by:
\begin{equation}
R_M / a = \sqrt{\frac{2U_{max}/\delta_{bond}}{6\pi \eta_0 a^2 \dot{\gamma}}} 
\label{R_M_definition}
\end{equation}
which corresponds to $Mn^{-0.5}$. Therefore, the $Mn^{-0.5}$ scaling shown in our simulations is consistent with a mechanical yielding criterion where the separating shear force balances the interparticle bond force.

\begin{figure*}[tbhp]
\centering
\includegraphics[width=\linewidth]{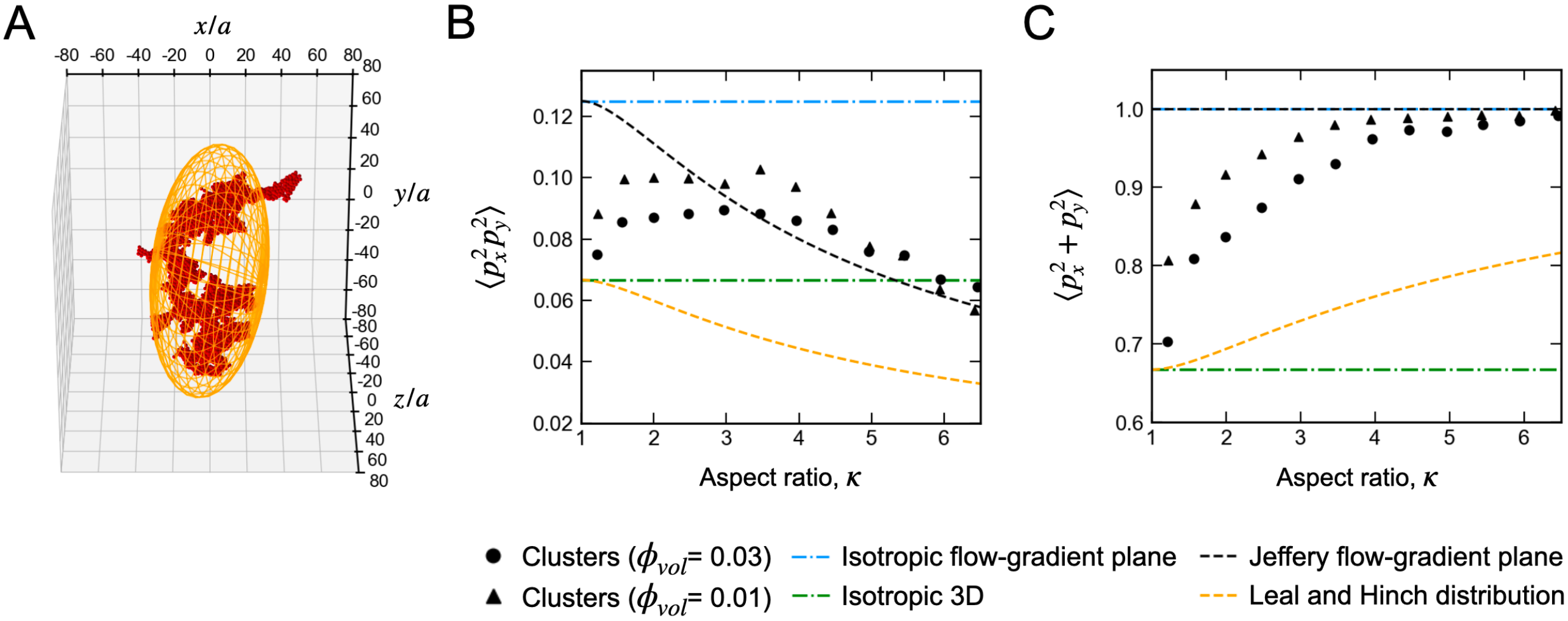}
\caption{Shape and orientation of clusters in the $R_g^2$ scaling regime. (A) Representative particle cluster and its equivalent ellipsoid at $Mn = 0.01$ and $\phi_{\mathit{vol}} = 0.03$. The principal semi-axes, $l_1$, $l_2$, and $l_3$, of the ellipsoid are proportional to the square roots of the three eigenvalues of the cluster gyration tensor. The eigenvector corresponding to the longest axis $l_1$ defines the cluster orientation $\mathbf{p}$. (B) Orientation average $\langle p_x^2 p_y^2 \rangle$ and (C) $\langle p_x^2 + p_y^2 \rangle$ as functions of the cluster aspect ratio $\kappa = l_1 \big/\sqrt{(l_2^2+l_3^2)/2}$. Symbols denote simulation results for $\phi_{\mathit{vol}} = 0.03$ and $0.01$. Dash-dotted lines indicate isotropic orientations in the flow-gradient plane (blue) and in 3D (green). Dashed lines denote Jeffery’s rotational motion of spheroids: the black line represents tumbling in the flow-gradient plane\ \cite{Jeffery1922motion}, while the orange line represents the Leal-Hinch orbit distribution for weak Brownian rotation\ \cite{leal1971effect}.}
\label{fig:3}
\end{figure*}

Fig.~\ref{fig:2} shows the relationship between the cluster stresslet $S_{cl,xy}$ and the radius of gyration $R_g$, scaled by $6\pi \eta_0 R_M^3 \dot{\gamma}$ and $R_M$, respectively. We note that the scaled cluster radius $R_g/R_M$ can exceed $1$. $R_M$ represents a single-bond failure criterion, which applies only in the dilute limit, where rare collisions prevent the formation of denser structures that could reinforce clusters against breakage. As seen in the comparison of $\phi_{\mathit{vol}} = 0.03$ with $\phi_{\mathit{vol}} = 0.01$, more frequent collisions drive this reinforcement, allowing clusters to exist well beyond $R_M$. In Fig.~\ref{fig:2}, all data collapse onto a master curve for all values of $\mathit{Mn}$. This master curve exhibits three distinct scaling regimes. At small cluster sizes, the stresslet increases as $R_g^3$, followed by an intermediate $R_g^2$ regime, before transitioning to an $R_g$ scaling at the highest values of $R_g$. The transition from $R_g^3$ to $R_g^2$ begins at approximately $R_g/R_M = 1$, suggesting that mechanical yielding plays an important role in the transition from $R_g^3$ to $R_g^2$ and eventually to $R_g$. To highlight this transition, the inset shows the cluster stresslet scaled by $6\pi \eta_0 R_M R_g^2 \dot{\gamma}$. Under this normalization, the $R_g^3$, $R_g^2$, and $R_g$ regimes appear as a linear increase, a plateau, and an inverse decay with $R_g/R_M$, respectively. At higher volume fraction and lower $\mathit{Mn}$, the plateau extends to larger $R_g/R_M$. This extension suggests that the $R_g^2$ regime represents a distinct cluster response rather than a smooth crossover between the $R_g^3$ and $R_g$ regimes. This response may reflect the interplay between hydrodynamic coupling and structural evolution as clusters become larger and denser. To investigate its origin, we examine the shape and orientational dynamics of individual clusters.

We have said that for rigid clusters, $S_{cl,xy}$ scales as $R_g^3$, which is consistent with the $R_g^3$ scaling at low $R_g$ in Fig.~\ref{fig:2}. Conversely, at the high-$R_g$ limit, $S_{cl,xy}$ is constrained by the maximum mechanical loading throughout the cluster over which the cluster breaks. In this regime, the cluster stresslet becomes $S_{cl,T} = F^{A}_{max} \times R_g = (2U_{max}/\delta_{bond}) R_g.$ Scaling the cluster stresslet by the hydrodynamic stresslet of a spherical cluster with the mechanical yielding radius $R_M$ gives
\begin{equation}
\frac{S_{cl,T}}{6\pi \eta_0 R_M^3 \dot{\gamma}} = \frac{2U_{max}/\delta_{bond}}{6\pi \eta_0 a^2 \dot{\gamma}} \left(\frac{a}{R_M}\right)^2 \frac{R_g}{R_M} = \frac{R_g}{R_M}. \label{Scl,T_vs_Rg/RM}
\end{equation}
Thus, the scaled $S_{cl,xy}$ for different $\mathit{Mn}$ collapses onto the unified curve at high $R_g$ when plotted as a function of $R_g/R_M$, while exhibiting its linear proportionality to $R_g$.

The underlying mechanism in the $R_g^2$ regime remains unclear. To address this, we examine the detailed cluster morphology and dynamics within this regime. Fig.~\ref{fig:3}A shows a representative cluster together with its equivalent ellipsoid. The orientations and lengths of the three principal axes of the ellipsoid are determined from the eigenvectors and eigenvalues, respectively, of the cluster gyration tensor $\mathbf{G} = \frac{1}{N}\sum_i \left(\mathbf{r}_i - \mathbf{r}_c\right) \left(\mathbf{r}_i - \mathbf{r}_c\right)$, where $\mathbf{r}_c$ is the cluster center of mass and $\mathbf{r}_i$ is the position of particle $i$ within the cluster. The eigenvalues of $\mathbf{G}$, denoted by $\lambda_1\geq\lambda_2\geq\lambda_3$, represent the squared
radii of gyration along the principal axes. For a uniform solid ellipsoid with semi-principal axes $l_1\geq l_2\geq l_3$, the squared radius of gyration satisfies $R_g^2=(l_1^2+l_2^2+l_3^2)/5=
\lambda_1+\lambda_2+\lambda_3$. Thus, the semi-principal axes of the ellipsoid are given by $l_i = \sqrt{5\lambda_i}$.

We characterize cluster shape using the aspect ratio $\kappa = l_1/\sqrt{(l_2^2+l_3^2)/2}$, and define the cluster orientation $\mathbf{p}$ as the direction of the principal eigenvector corresponding to the largest eigenvalue $l_1$. In Fig.~\ref{fig:3}B and C, we plot the orientation averages $\langle p_x^2 p_y^2 \rangle$ and $\langle p_x^2 + p_y^2 \rangle$ as a function of $\kappa$ for $\phi_{\mathit{vol}} = 0.03$ and $0.01$.

We compare these results to four orientation distribution scenarios: the isotropic flow-gradient plane, isotropic 3D, the Jeffery flow-gradient plane\ \cite{Jeffery1922motion}, and the Leal and Hinch distribution for weak Brownian rotation\ \cite{leal1971effect}. The Jeffery flow-gradient plane distribution and Leal and Hinch distribution shown here are derived for rigid spheroids. The general equations of rotational motion for spheroids in simple shear flow, originally derived by Jeffery\ \cite{Jeffery1922motion}, are given by $\dot{\phi}=\frac{\dot{\gamma}}{\kappa^2 + 1} \left( \kappa^2 \sin^2 \phi + \cos^2 \phi \right)$ and $\dot{\theta}=\frac{\dot{\gamma} (\kappa^2 - 1)}{4 (\kappa^2 + 1)} \sin 2\phi \sin 2\theta$
where $\phi$ is the angle of the spheroid’s orientation projected onto the flow-gradient ($x$-$y$) plane relative to the flow ($x$-) direction and $\theta$ is the polar angle relative to the vorticity ($z$-) direction. Integrating these equations gives:

\begin{equation}
\begin{aligned}
\tan(\phi) &= \kappa \tan\left(\frac{\dot{\gamma} t}{\kappa + 1/\kappa}\right) \\ 
\tan(\theta) &= \frac{C \kappa}{(\kappa^2 \sin^2 \phi + \cos^2 \phi )^{0.5}} 
\label{angle_vs_time}
\end{aligned}
\end{equation}
where $C$ is the orbit constant. Using Eq.~\ref{angle_vs_time}, the orientation vector $\mathbf{p} = \sin \theta \cos \phi \hat{\mathbf{e}}_x + \sin \theta \sin \phi \hat{\mathbf{e}}_y + \cos \theta \hat{\mathbf{e}}_z$ can be expressed in terms of $\kappa$ and $C$. The time-averaged orientations $\langle p_x^2 p_y^2 \rangle_C$ and $\langle p_x^2 + p_y^2 \rangle_C$ can then be obtained for a given orbit $C$. Leal and Hinch analytically determined the orbit distributions $f(C)$ for spheroids undergoing weak Brownian motion\ \cite{leal1971effect, rahnama1995orbit}. Integrating the orientation averages for a given $C$ over this distribution $f(C)$ provides the ensemble averages shown in Fig.~\ref{fig:3}B and C.

\begin{figure}[t!]
\centering
\includegraphics[width=\linewidth]{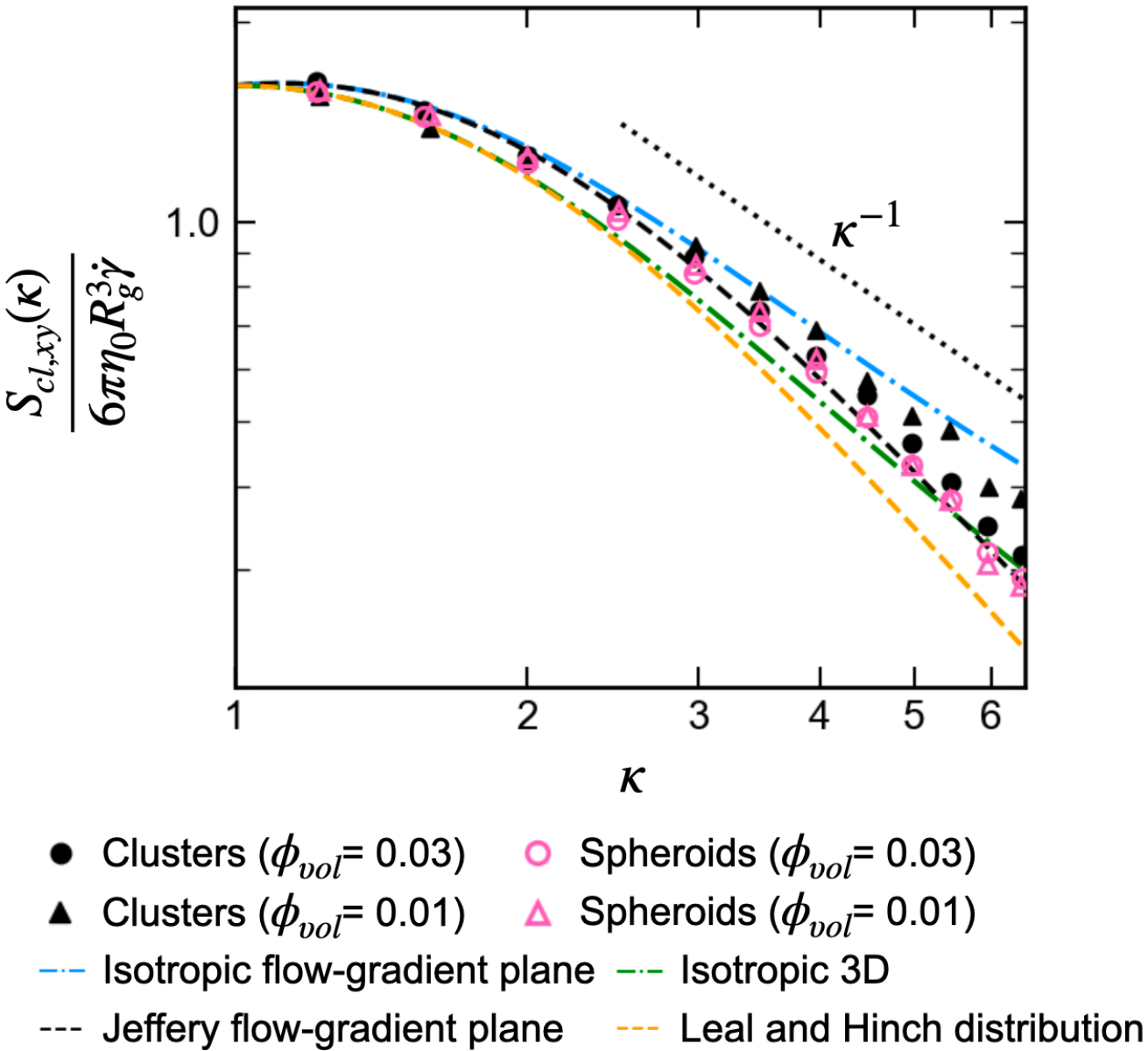}
\caption{Effect of cluster shape and orientation on the stresslet in the $R_g^2$ scaling regime. The cluster stresslet $S_{\mathit{cl,xy}}$, normalized by the rigid-sphere scaling $6 \pi \eta_0 R_g^3 \dot{\gamma}$, is plotted as a function of the cluster aspect ratio $\kappa$. Symbols denote simulation results for clusters at $\phi_{\mathit{vol}} = 0.03$ and $0.01$. Open symbols represent equivalent spheroids having the same orientation averages as those obtained in the simulations, with stresslets computed using the theoretical hydrodynamic resistance functions for spheroids. Dash-dotted lines indicate isotropic orientation distributions in the flow-gradient plane (blue) and in 3D (green). Dashed lines denote theoretical values for Jeffery rotation in the flow-gradient plane (black) and the Leal and Hinch distribution for weak Brownian rotation (orange). The dotted guideline indicates a $\kappa^{-1}$ scaling.}
\label{fig:4}
\end{figure}

Fig.~\ref{fig:3}B and C reveal the evolution of cluster orientation with increasing $\kappa$. For nearly spherical clusters ($\kappa \approx 1$), both $\langle p_x^2 p_y^2 \rangle$ and $\langle p_x^2 + p_y^2 \rangle$ are close to the isotropic 3D and Leal and Hinch distributions. As $\kappa$ increases, however, both orientation averages rapidly approach the Jeffery flow-gradient plane distribution ($C = \infty$). This suggests that collisions occur preferentially within the flow-gradient plane, and that these repeated processes drive clusters to form anisotropic shapes and become aligned within that plane\ \cite{iso1996orientation, sundararajakumar1997orientation, petrich2000orientation}. For $\kappa > 3$, $\langle p_x^2 p_y^2 \rangle$ decreases and closely follows the Jeffery flow-gradient plane distribution. This decrease is attributed to the alignment of spheroids with the flow direction; the rotational velocity of a spheroid is $\kappa^2$ times slower when it is aligned with the flow direction than when aligned with the gradient direction. High-$\kappa$ clusters spend a larger fraction of their rotation period aligned with the flow direction, reducing the time-averaged value of $\langle p_x^2 p_y^2 \rangle$.

We next examine how Jeffery’s rotational dynamics identified in Fig.~\ref{fig:3} modify the hydrodynamic stresslet of clusters. For a rigid spheroid in simple shear flow, Jeffery derived the stresslet as $S_{cl,xy} = \eta_0 a^3 \dot{\gamma} \left( A\langle p_x^2 p_y^2 \rangle + B\langle p_x^2 + p_y^2 \rangle + C \right)$, where the coefficients $A$, $B$, and $C$ depend only on the aspect ratio $\kappa$\ \cite{Jeffery1922motion, batchelor1970stress, karrila1992microhydrodynamics}. Fig.~\ref{fig:4} shows the cluster stresslet normalized by the rigid-sphere scaling $6 \pi \eta_0 R_g^3 \dot{\gamma}$ as a function of $\kappa$. For nearly spherical clusters, the simulation results agree well with the theoretical values for isotropic 3D and Leal and Hinch distributions. As $\kappa$ increases, however, the normalized stresslet gradually approaches the Jeffery flow-gradient plane curve. This behavior is consistent with the orientational dynamics observed in Fig.~\ref{fig:3}B and C, confirming that cluster collisions occur within the flow-gradient plane, resulting in 2D Jeffery rotational motion dominating the ensemble-averaged stresslet.

\begin{figure}[t!]
\centering
\includegraphics[width=\linewidth]{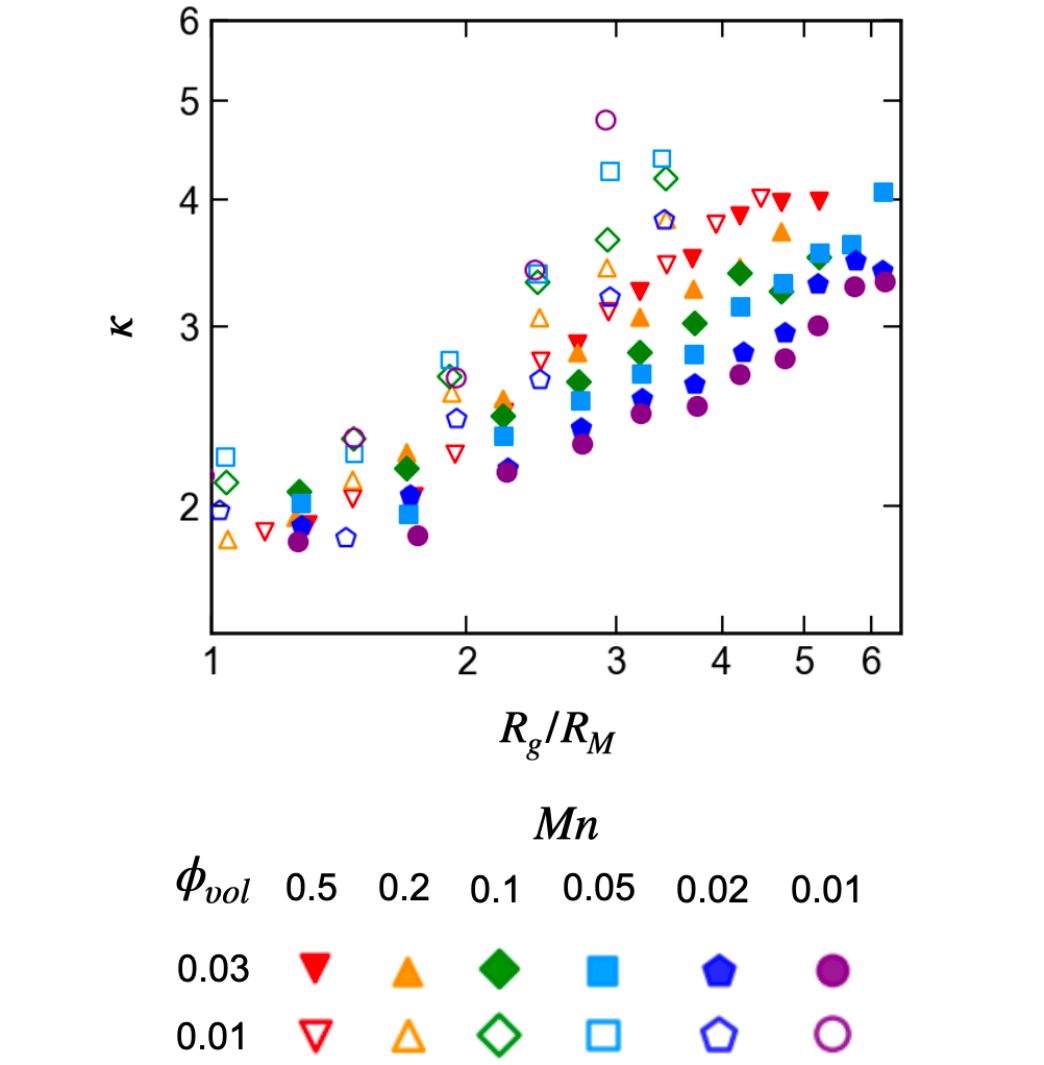}
\caption{Average cluster aspect ratio $\kappa$ as a function of the scaled cluster radius $R_g/R_M$ for $\phi_{\mathit{vol}}=0.01$ and $0.03$. The radius of gyration $R_g$ is scaled by the mechanical yielding radius $R_M$ to connect the evolution of cluster shape to the unified stresslet scaling established in Fig. 2.}
\label{fig:5}
\end{figure}

The stresslets computed for equivalent spheroids using the cluster orientation averages obtained in Fig. 3B and C agree well with the simulation results for $\kappa < 3$. For larger aspect ratios, however, the spheroidal stresslet underestimates the cluster stresslet. This discrepancy indicates that the cluster stresslet is influenced not only by the cluster orientation but also by deviations of the cluster morphology from an ideal spheroid. As visualized in Fig.~\ref{fig:3}A, real clusters possess rough, concave surfaces, as well as kinks protruding from the equivalent spheroidal surfaces. These non-uniform surfaces induce larger hydrodynamic stresses than a smooth spheroidal surface of the same $\kappa$, inducing a larger cluster stresslet\ \cite{meng2008large, singh2011ideal}.

At large aspect ratios, both the simulation results and the Jeffery flow-gradient plane prediction decay approximately as $\kappa^{-1}$ (Fig.~\ref{fig:4}). This decay arises from the cluster orientation dynamics established in Fig.~\ref{fig:3}. Clusters with high $\kappa$ spend a large fraction of their rotational period aligned with the flow direction, where the disturbance to the ambient flow and the resulting induced stresslet are minimized. The dominant contribution to the average stresslet instead arises during the rapid flipping phase, producing a stresslet scaling as $R_g^3$. Because the fraction of the rotational period spent in this flipping phase scales as $\kappa^{-1}$, averaging over the rotational period gives a cluster stresslet that scales as $R_g^3/\kappa$.

\begin{figure}[t!]
\centering
\includegraphics[width=\linewidth]{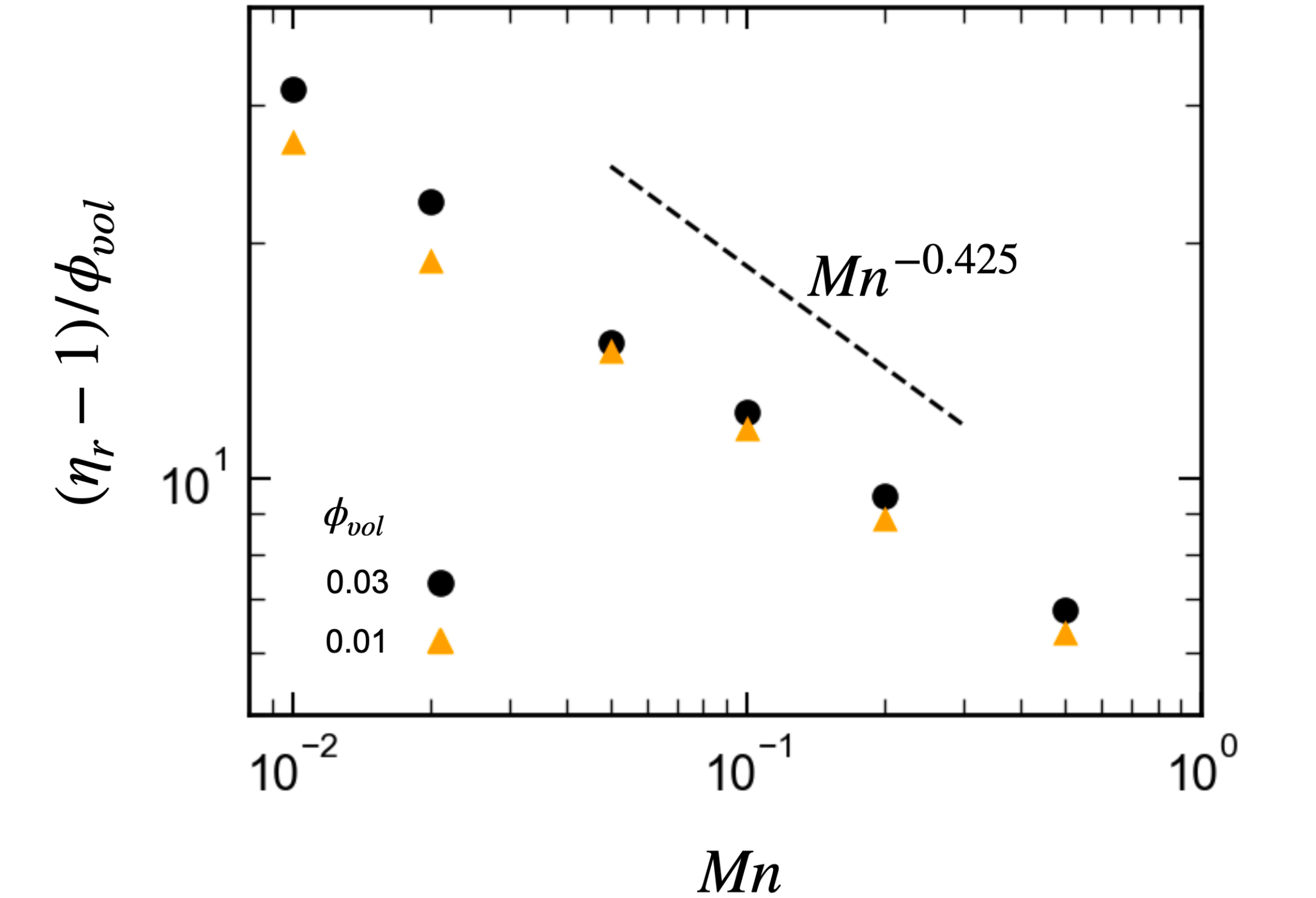}
\caption{Particle contribution to the relative viscosity normalized by volume fraction, $(\eta_r-1)/\phi_{\mathit{vol}}$, as a function of $\mathit{Mn}$. The dashed line indicates an $\mathit{Mn}^{-0.425}$ scaling.}
\label{fig:6}
\end{figure}

Fig.~\ref{fig:5} shows the average cluster aspect ratio $\kappa$ as a function of the scaled cluster radius $R_g/R_M$ for different $\mathit{Mn}$ and $\phi_{\mathit{vol}}$. For $R_g/R_M < 1.5$, $\kappa$ remains nearly constant at $\kappa \approx 2$. At larger cluster sizes, however, $\kappa$ increases continuously with $R_g/R_M$. We note that this increase arises from shear-induced mechanical yielding. As shown in Fig.~\ref{fig:4}, clusters with lower $\kappa$ induce larger stresslets than high-$\kappa$ clusters of the same $R_g$. These low-$\kappa$ clusters reach their critical mechanical yielding stress and yield first. The preferential rupture of these low-$\kappa$ clusters increases the average aspect ratio of the remaining cluster population. As $R_g$ grows, clusters with higher $\kappa$ induce stresslets large enough to trigger mechanical yielding, leading to a continuous increase in the average aspect ratio with cluster size.

Fig.~\ref{fig:6} shows the particle contribution to the suspension relative viscosity. Its dependence on $\mathit{Mn}$ is consistent with an $\mathit{Mn}^{-0.425}$ scaling at both volume fractions. This can be understood as follows: The particle number density scales as $n\sim\phi_{\mathit{vol}}/a^3$, while the number of particles within a cluster scales as $N_g\sim(R_g/a)^{d_f}$, where $d_f$ is the cluster fractal dimension, giving a cluster number density that scales as $n_c\sim n/N_g\sim n(R_g/a)^{-d_f}$. Using the cluster stresslet scaling, $S_{cl,xy}\sim\eta_0\dot{\gamma}R_M^{3-x}R_g^x$, where $1\le x\le3$ according to the $R_g/R_M$ range (Fig.~\ref{fig:2}), gives
\begin{equation}
\frac{\eta_r-1}{\phi_{\mathit{vol}}}
\sim\frac{n_c S_{cl,xy}}
{n a^3\eta_0\dot{\gamma}}
\sim\left(\frac{R_M}{a}\right)^{3-d_f}
\left(\frac{R_g}{R_M}\right)^{x-d_f}.
\label{rel_vis}
\end{equation}
Using the average fractal dimension $d_f=2.15$ obtained from the simulations gives an $\mathit{Mn}^{-0.425}$ scaling.

\section*{Discussion}
\noindent Our results reveal the physical origins of the $R_g^3$, $R_g^2$, and $R_g$ scaling regimes of the cluster stresslet formed in sheared attractive suspensions. The $R_g^3$ regime reflects the hydrodynamic stress produced by small, rigid clusters, while the $R_g$ regime arises from the constraint imposed by maximum interparticle bond loading. The $R_g^2$ regime can be explained by the coupled evolution of cluster shape and orientation under shear-induced mechanical yielding. In this regime, clusters with low aspect ratios rupture preferentially, driving an increase in the average aspect ratio of the remaining cluster population. These clusters exhibit Jeffery rotational motion in the flow-gradient plane, inducing an average stresslet that scales as $R_g^3/\kappa$. Combining these relationships produces the $R_g^2$ scaling. Our results connect microscopic interparticle interactions and hydrodynamics to cluster morphology, rotational dynamics, and macroscopic suspension stress. These scaling relations provide a framework for selecting shear conditions to control cluster structure and rheology in attractive suspensions, with potential applications in processes requiring engineered material properties.

\clearpage
\onecolumngrid
\def\bibsection{} % Suppresses the heading inside \bibliography
\section*{References}
\vspace{-0.0em} % Adjust negative spacing as needed
% Bibliography
\bibliography{pnas-reference}

@article{papadopoulou2020shear,
  title={On the shear thinning of non-Brownian suspensions: Friction or adhesion?},
  author={Papadopoulou, Anastasia and Gillissen, Jurriaan J and Wilson, Helen J and Tiwari, Manish K and Balabani, Stavroula},
  journal={Journal of Non-Newtonian Fluid Mechanics},
  volume={281},
  pages={104298},
  year={2020},
  publisher={Elsevier}
}

@article{hipp2024ElecTransport,
  title={Quantifying electron transport in aggregated colloidal suspensions in the strong flow regime},
  author={Hipp, Julie B and Ramos, Paolo Z and Liu, Qingsong and Wagner, Norman J and Richards, Jeffrey J},
  journal={Proceedings of the National Academy of Sciences},
  volume={121},
  number={34},
  pages={e2403000121},
  year={2024},
  publisher={National Academy of Sciences}
}

@article{li2013SSFB,
  title={Aqueous semi-solid flow cell: demonstration and analysis},
  author={Li, Zheng and Smith, Kyle C and Dong, Yajie and Baram, Nir and Fan, Frank Y and Xie, Jing and Limthongkul, Pimpa and Carter, W Craig and Chiang, Yet-Ming},
  journal={Physical Chemistry Chemical Physics},
  volume={15},
  number={38},
  pages={15833--15839},
  year={2013},
  publisher={The Royal Society of Chemistry}
}

@article{das2024surfactantACparticles,
  title={Surfactant-driven dynamic changes in rheology of activated carbon slurry electrodes},
  author={Das, Mohan and Lee, KangJin and Wirth, Christopher L},
  journal={ACS Applied Materials \& Interfaces},
  volume={16},
  number={32},
  pages={42049--42058},
  year={2024},
  publisher={ACS Publications}
}

@article{jin2025ACparticles,
  title={Carbon particle aggregation for enhanced flow capacitive deionization},
  author={Jin, Ganggang and Zhu, Qinpeng and Guo, Jiayue and Zhang, Liuyang and Xu, Xiangyang and Yang, Peihua},
  journal={Chemical Communications},
  volume={61},
  number={80},
  pages={15610--15613},
  year={2025},
  publisher={The Royal Society of Chemistry}
}

@article{biggs2000PolymerBridging,
  title={Aggregate structures formed via a bridging flocculation mechanism},
  author={Biggs, Simon and Habgood, Michael and Jameson, Graeme J and Yan, Yao-de},
  journal={Chemical Engineering Journal},
  volume={80},
  number={1-3},
  pages={13--22},
  year={2000},
  publisher={Elsevier}
}

@article{agarwal2009PolymerBridgingModel,
  title={A model for the collision efficiency of shear-induced agglomeration involving polymer bridging},
  author={Agarwal, Sushant and Gupta, Rakesh K and Doraiswamy, Deepak},
  journal={Colloids and Surfaces A: Physicochemical and Engineering Aspects},
  volume={345},
  number={1-3},
  pages={224--230},
  year={2009},
  publisher={Elsevier}
}

@article{Zaccone2011BrowntoNonBrown,
  title = {Shear-Driven Solidification of Dilute Colloidal Suspensions},
  author = {Zaccone, Alessio and Gentili, Daniele and Wu, Hua and Morbidelli, Massimo and Del Gado, Emanuela},
  journal = {Phys. Rev. Lett.},
  volume = {106},
  issue = {13},
  pages = {138301},
  numpages = {4},
  year = {2011},
  month = {Mar},
  publisher = {American Physical Society},
  doi = {10.1103/PhysRevLett.106.138301},
  url = {https://link.aps.org/doi/10.1103/PhysRevLett.106.138301}
}

@article{pongjanyakul2009Pharma,
  title={Polymer--magnesium aluminum silicate composite dispersions for improved physical stability of acetaminophen suspensions},
  author={Pongjanyakul, Thaned and Puttipipatkhachorn, Satit},
  journal={AAPS PharmSciTech},
  volume={10},
  number={2},
  pages={346--354},
  year={2009},
  publisher={Springer}
}

@article{lin2022diffusivity,
  title={Quantifying the hydrodynamic contribution to electrical transport in non-Brownian suspensions},
  author={Lin, Han and Majji, Madhu V and Cho, Noah and Zeeman, John R and Swan, James W and Richards, Jeffrey J},
  journal={Proceedings of the National Academy of Sciences},
  volume={119},
  number={29},
  pages={e2203470119},
  year={2022},
  publisher={National Academy of Sciences}
}

@article{swan2016rapid,
  title={Rapid calculation of hydrodynamic and transport properties in concentrated solutions of colloidal particles and macromolecules},
  author={Swan, James W and Wang, Gang},
  journal={Physics of Fluids},
  volume={28},
  number={1},
  year={2016},
  publisher={AIP Publishing}
}

@article{VargaSwan2019PatternFormation,
 author={Zsigmond Varga  and Vincent Grenard  and Stefano Pecorario  and Nicolas Taberlet  and Vincent Dolique  and Sébastien Manneville  and Thibaut Divoux  and Gareth H. McKinley  and James W. Swan},
 title = {Hydrodynamics control shear-induced pattern formation in attractive suspensions},
 journal = {Proceedings of the National Academy of Sciences},
 volume = {116},
 number = {25},
 pages = {12193-12198},
 year = {2019}
}

@article{manounou2014discrete,
  title={Discrete element modeling of the microstructure of fine particle agglomerates in sheared dilute suspension},
  author={Manounou, A Kimbonguila and R{\'e}mond, S},
  journal={Physica A: Statistical Mechanics and its Applications},
  volume={412},
  pages={66--83},
  year={2014},
  publisher={Elsevier}
}

@article{Irani2014AggregatingDisks,
  title = {Impact of Attractive Interactions on the Rheology of Dense Athermal Particles},
  author = {Irani, Ehsan and Chaudhuri, Pinaki and Heussinger, Claus},
  journal = {Phys. Rev. Lett.},
  volume = {112},
  issue = {18},
  pages = {188303},
  numpages = {5},
  year = {2014},
  month = {May},
  publisher = {American Physical Society},
  doi = {10.1103/PhysRevLett.112.188303},
  url = {https://link.aps.org/doi/10.1103/PhysRevLett.112.188303}
}

@article{Singh2019YiedlingtoJamming,
  title = {From Yielding to Shear Jamming in a Cohesive Frictional Suspension},
  author = {Singh, Abhinendra and Pednekar, Sidhant and Chun, Jaehun and Denn, Morton M. and Morris, Jeffrey F.},
  journal = {Phys. Rev. Lett.},
  volume = {122},
  issue = {9},
  pages = {098004},
  numpages = {5},
  year = {2019},
  month = {Mar},
  publisher = {American Physical Society},
  doi = {10.1103/PhysRevLett.122.098004},
  url = {https://link.aps.org/doi/10.1103/PhysRevLett.122.098004}
}

@article{cichocki1999ThreeBody,
  title={Lubrication corrections for three-particle contribution to short-time self-diffusion coefficients in colloidal dispersions},
  author={Cichocki, B and Ekiel-Je{\.z}ewska, ML and Wajnryb, E},
  journal={The Journal of chemical physics},
  volume={111},
  number={7},
  pages={3265--3273},
  year={1999},
  publisher={American Institute of Physics}
}

@article{abade2010HighFreqVisc,
  title={High-frequency viscosity of concentrated porous particles suspensions},
  author={Abade, Gustavo C and Cichocki, Bogdan and Ekiel-Je{\.z}ewska, Maria L and N{\"a}gele, Gerhard and Wajnryb, Eligiusz},
  journal={The Journal of chemical physics},
  volume={133},
  number={8},
  year={2010},
  publisher={AIP Publishing}
}

@article{Zia2017pairMobility,
  title={Pair mobility functions for rigid spheres in concentrated colloidal dispersions: stresslet and straining motion couplings},
  author={Su, Yu and Swan, James W and Zia, Roseanna N},
  journal={The Journal of chemical physics},
  volume={146},
  number={12},
  year={2017},
  publisher={AIP Publishing}
}

@article{duduta2011SSFB,
  title={Semi-solid lithium rechargeable flow battery},
  author={Duduta, Mihai and Ho, Bryan and Wood, Vanessa C and Limthongkul, Pimpa and Brunini, Victor E and Carter, W Craig and Chiang, Yet-Ming},
  journal={Advanced Energy Materials},
  volume={1},
  number={4},
  pages={511--516},
  year={2011},
  publisher={WILEY-VCH Verlag Weinheim}
}

@article{brunini2012modelingSSFB,
  title={Modeling the hydrodynamic and electrochemical efficiency of semi-solid flow batteries},
  author={Brunini, Victor E and Chiang, Yet-Ming and Carter, W Craig},
  journal={Electrochimica acta},
  volume={69},
  pages={301--307},
  year={2012},
  publisher={Elsevier}
}

@article{fan2014polysulfideSSFB,
  title={Polysulfide flow batteries enabled by percolating nanoscale conductor networks},
  author={Fan, Frank Y and Woodford, William H and Li, Zheng and Baram, Nir and Smith, Kyle C and Helal, Ahmed and McKinley, Gareth H and Carter, W Craig and Chiang, Yet-Ming},
  journal={Nano letters},
  volume={14},
  number={4},
  pages={2210--2218},
  year={2014},
  publisher={ACS Publications}
}

@article{fiore2018rapid,
  title={Rapid sampling of stochastic displacements in Brownian dynamics simulations with stresslet constraints},
  author={Fiore, Andrew M and Swan, James W},
  journal={The Journal of chemical physics},
  volume={148},
  number={4},
  year={2018},
  publisher={AIP Publishing}
}

@article{fiore2019FSD,
  title={Fast stokesian dynamics},
  author={Fiore, Andrew M and Swan, James W},
  journal={Journal of Fluid Mechanics},
  volume={878},
  pages={544--597},
  year={2019},
  publisher={Cambridge University Press}
}

@article{brady1988stokesian,
  title={Stokesian dynamics},
  author={Brady, John F and Bossis, Georges},
  journal={Annual Review of Fluid Mechanics}, 
  volume={20},
  pages={111--157},
  year={1988}
}

@article{sierou2001accelerated,
  title={Accelerated Stokesian dynamics simulations},
  author={Sierou, Asimina and Brady, John F},
  journal={Journal of fluid mechanics},
  volume={448},
  pages={115--146},
  year={2001},
  publisher={Cambridge University Press}
}

@article{wang2016spectral,
  title={Spectral Ewald acceleration of Stokesian dynamics for polydisperse suspensions},
  author={Wang, Mu and Brady, John F},
  journal={Journal of Computational Physics},
  volume={306},
  pages={443--477},
  year={2016},
  publisher={Elsevier}
}

@article{varga2018large,
  title={Large scale anisotropies in sheared colloidal gels},
  author={Varga, Zsigmond and Swan, James W},
  journal={Journal of Rheology},
  volume={62},
  number={2},
  pages={405--418},
  year={2018},
  publisher={AIP Publishing}
}

@book{andreotti2013granular,
  title={Granular media: between fluid and solid},
  author={Andreotti, Bruno and Forterre, Yo{\"e}l and Pouliquen, Olivier},
  year={2013},
  publisher={Cambridge University Press}
}

@article{gallier2014Friction,
  title={Rheology of sheared suspensions of rough frictional particles},
  author={Gallier, Stany and Lemaire, Elisabeth and Peters, Fran{\c{c}}ois and Lobry, Laurent},
  journal={Journal of Fluid Mechanics},
  volume={757},
  pages={514--549},
  year={2014},
  publisher={Cambridge University Press}
}

@article{mari2015DST,
  title={Discontinuous shear thickening in Brownian suspensions by dynamic simulation},
  author={Mari, Romain and Seto, Ryohei and Morris, Jeffrey F and Denn, Morton M},
  journal={Proceedings of the National Academy of Sciences},
  volume={112},
  number={50},
  pages={15326--15330},
  year={2015},
  publisher={National Academy of Sciences}
}

@article{zhang2024friction,
  title={Frictional effects on shear-induced diffusion in suspensions of non-Brownian particles},
  author={Zhang, Han and Kopelevich, Dmitry I and Butler, Jason E},
  journal={Journal of Fluid Mechanics},
  volume={1001},
  pages={A42},
  year={2024},
  publisher={Cambridge University Press}
}

@article{fernandez2013Friction,
  title={Microscopic mechanism for shear thickening of non-Brownian suspensions},
  author={Fernandez, Nicolas and Mani, Roman and Rinaldi, David and Kadau, Dirk and Mosquet, Martin and Lombois-Burger, H{\'e}lene and Cayer-Barrioz, Juliette and Herrmann, Hans J and Spencer, Nicholas D and Isa, Lucio},
  journal={Physical review letters},
  volume={111},
  number={10},
  pages={108301},
  year={2013},
  publisher={APS}
}

@article{da1996CunhaHinch,
  title={Shear-induced dispersion in a dilute suspension of rough spheres},
  author={Da Cunha, FR and Hinch, EJ},
  journal={Journal of fluid mechanics},
  volume={309},
  pages={211--223},
  year={1996},
  publisher={Cambridge University Press}
}

@article{potanin1993aggregate,
  title={On the computer simulation of the deformation and breakup of colloidal aggregates in shear flow},
  author={Potanin, Andrew A},
  journal={Journal of colloid and interface science},
  volume={157},
  number={2},
  pages={399--410},
  year={1993},
  publisher={Elsevier}
}

@article{Jeffery1922motion,
  title={The motion of ellipsoidal particles immersed in a viscous fluid},
  author={Jeffery, George Barker},
  journal={Proceedings of the Royal Society of London. Series A, Containing papers of a mathematical and physical character},
  volume={102},
  number={715},
  pages={161--179},
  year={1922},
  publisher={The Royal Society London}
}

@article{leal1971effect,
  title={The effect of weak Brownian rotations on particles in shear flow},
  author={Leal, LG and Hinch, EJ},
  journal={Journal of Fluid Mechanics},
  volume={46},
  number={4},
  pages={685--703},
  year={1971},
  publisher={Cambridge University Press}
}

@article{rahnama1995orbit,
  title={The effect of hydrodynamic interactions on the orientation distribution in a fiber suspension subject to simple shear flow},
  author={Rahnama, Mani and Koch, Donald L and Shaqfeh, Eric SG},
  journal={Physics of Fluids},
  volume={7},
  number={3},
  pages={487--506},
  year={1995},
  publisher={American Institute of Physics}
}

@article{iso1996orientation,
  title={Orientation in simple shear flow of semi-dilute fiber suspensions 1. Weakly elastic fluids},
  author={Iso, Yoichi and Koch, Donald L and Cohen, Claude},
  journal={Journal of non-newtonian fluid mechanics},
  volume={62},
  number={2-3},
  pages={115--134},
  year={1996},
  publisher={Elsevier}
}

@article{sundararajakumar1997orientation,
  title={Structure and properties of sheared fiber suspensions with mechanical contacts},
  author={Sundararajakumar, RR and Koch, Donald L},
  journal={Journal of Non-Newtonian Fluid Mechanics},
  volume={73},
  number={3},
  pages={205--239},
  year={1997},
  publisher={Elsevier}
}

@article{petrich2000orientation,
  title={An experimental determination of the stress--microstructure relationship in semi-concentrated fiber suspensions},
  author={Petrich, Michael P and Koch, Donald L and Cohen, Claude},
  journal={Journal of non-newtonian fluid mechanics},
  volume={95},
  number={2-3},
  pages={101--133},
  year={2000},
  publisher={Elsevier}
}

@article{batchelor1970stress,
  title={The stress system in a suspension of force-free particles},
  author={Batchelor, GK},
  journal={Journal of fluid mechanics},
  volume={41},
  number={3},
  pages={545--570},
  year={1970},
  publisher={Cambridge University Press}
}

@article{karrila1992microhydrodynamics,
  title={Microhydrodynamics: principles and selected applications},
  author={Karrila, Seppo},
  journal={Applied Mathematical Modelling},
  year={1992}
}

@article{meng2008large,
  title={Large scale dynamic simulation of plate-like particle suspensions. Part I: Non-Brownian simulation},
  author={Meng, Qingjun and Higdon, Jonathan JL},
  journal={Journal of Rheology},
  volume={52},
  number={1},
  pages={1--36},
  year={2008},
  publisher={AIP Publishing}
}

@article{singh2011ideal,
  title={Ideal rate of collision of cylinders in simple shear flow},
  author={Singh, Vikram and Koch, Donald L and Stroock, Abraham D},
  journal={Langmuir},
  volume={27},
  number={19},
  pages={11813--11823},
  year={2011},
  publisher={ACS Publications}
}

\end{document}